\documentclass[aps,prd,groupedaddress,superscriptaddress,11pt]{revtex4}

\usepackage{amsmath,amssymb,bm}
\usepackage{graphicx}
\usepackage{epstopdf}
\usepackage{amsfonts}
\usepackage{amssymb}
\usepackage{amsbsy}
\usepackage{amsmath}
\usepackage{latexsym}
\usepackage{sansmath}
\usepackage{bm}
\usepackage{color}
\usepackage{comment}
\usepackage{csquotes}								% for nice looking quotation marks.
\usepackage{physics}
\usepackage{eufrak}
\usepackage{accents}
\usepackage[colorlinks=true,breaklinks]{hyperref}

\usepackage[plain]{algorithm}
\usepackage{algpseudocode}

\def\be{\begin{equation}}
\def\ee{\end{equation}}
\def\bea {\begin{eqnarray}}
\def\eea {\end{eqnarray}}
\def\nn {\nonumber}
\def \p {\partial}

\begin{document}

\title{Semiclassical cosmology with back  reaction: \\the Friedmann-Schrodinger equation and inflation}
\author{Viqar Husain} \email{vhusain@unb.ca} 
\affiliation{Department of Mathematics and Statistics, University of New Brunswick, Fredericton, NB, Canada E3B 5A3}
\author{Suprit Singh} \email{suprit.singh@unb.ca} 
\affiliation{Department of Mathematics and Statistics, University of New Brunswick, Fredericton, NB, Canada E3B 5A3}

\begin{abstract}
\vskip 0.2cm

We propose and study a semi-classical cosmological system akin to the Newton-Schr\"odinger equation where matter field evolution is determined  by time dependent Schr\"odinger equation. The resulting dynamics is one where the scale factor self-consistently informs the quantum evolution of the scalar field wave function via the Friedmann equation.  We study various potentials, and  show that for each case,  inflation arises naturally, without fine-tuning, for arbitrary  initial wave functions for the scalar field.  Furthermore, due to zero-point quantum fluctuations of the scalar field, the equation of state $P=-\rho$ arises dynamically at late times for all initial states and potentials.  
\end{abstract}

\maketitle

\section{Introduction}

There is at present no widely accepted theory of quantum gravity. There is, however, a large volume of work on quantum fields propagating on a fixed curved spacetime. This work is of wide scope, addressing both the mathematical structure and physical effects pertaining to black holes and cosmology (See eg. \cite{Parker:2009uva} ). Quantum fields on curved spacetime is also often referred to as ``semiclassical gravity,"  because curved spacetimes imply a non-zero classical gravitational field, and contain the three constant $c$, $G$ and $\hbar$.  But working with quantum field theory on a fixed curved spacetime does not address the important question of back-reaction. This requires a formulation of semiclassical gravity  in which there is coupled dynamics of classical gravity and quantum matter. The prototype for this is the semi-classical Einstein equation,
\be
G_{ab}(g) + \Lambda g_{ab} = 8\pi G\  \langle \Psi | \hat{T}_{ab}(g,\hat{\phi}) | \Psi \rangle, \label{sc1}
\ee
where the ``semiclassical metric" $g$ is the unknown, and is to be solved for, given a state $|\Psi\rangle$ in the Hilbert space of the matter theory in which $\hat{\phi}$ is an elementary operator.  This equation has several issues, both in its precise formulation and in its interpretation \cite{Isham:1995wr}. Nevertheless to date it appears to be ``the only game in town"; all works we are aware of use some form of this equation. 

 In this paper we formulate a new way to couple classical gravity with quantum matter in the cosmological context. The main idea comes from noting that in the semiclassical equation (\ref{sc1}), all time derivatives of the metric $g$ come from the l.h.s., so the dynamics in this equation may be called ``gravitationally determined". Indeed, the r.h.s. is just a function of the metric after the expectation value is computed (provided that is possible without ambiguity for an arbitrary $g$). Therefore, there is no independent unitary quantum evolution of matter in this equation.   
  
The semi-classical gravity equation we describe here is different in that it involves introducing the time dependent Schr\"odinger equation (TDSE) for the matter wave function as a basic premise.  Therefore, it contains an additional dynamics which is separate from the time evolution due to Einstein equations. This is similar in spirit to the so-called Newton-Schr\"odinger equation \cite{Diosi:2014ura,Carlip:2008zf,Bahrami:2014gwa} where the Poisson equation for the gravitational potential $\phi$ is sourced by the matter wave function $|\Psi |^2$, which in turn satisfies the Schr\"odinger equation with a gravitational potential $\phi$.  While we do not propose a formulation that applies  generally, we give a precise  self-consistent version that applies to homogeneous and isotropic cosmology, as well as a  generalization to Bianchi models. This provides an approach for non-perturbative calculations that include  back-reaction.  

The formulation  we give has potential impact for questions such as the ``probability of inflation", which is usually addressed in the context of a purely classical dynamical system describing the evolution of a homogeneous matter field and the scale factor (-- see e.g. \cite{Ashtekar:2009mm, Ijjas:2014nta, Corichi:2013kua, Brandenberger:2016uzh} for recent discussions). In our approach the space of initial data is vastly enlarged to a subspace of the square-integrable functions -- the initial state vector of matter. This places the questions about inflation in an entirely different ``landscape" of data. Our main result is that for a variety of scalar potentials, there are uncountably large initial data sets of functions describing quantum states of matter that all lead to an inflationary Universe.  

Although we do not discuss inhomogeneities in this paper, our approach treats the zero modes of the scalar field quantum mechanically rather than classically (as in the standard treatment). This makes it consistent with the standard view of treating perturbations as quantum. It also provides a starting point for treating perturbations in a setting where both the zero and non-zero modes are treated in quantum theory. 
 
The outline of the paper is as follows: In the next Section we discuss the conventional semi-classical approximation and present our formulation. In Sec. III we describe the numerical method for finding solutions and present our results.   We conclude with a summary and discussion in Sec. IV.   The Appendix contains details of our numerical method.

\section{Semiclassical approximations}

The semiclassical approximation defined by (\ref{sc1}) has several conceptual and computational problems. As discussed in \cite{Isham:1995wr}, these include the following: 

\begin{itemize}

\item The expectation value on the r.h.s. cannot be computed unambiguously in any but the simplest backgrounds with high degrees of symmetry -- there is no reason for semi-classical metrics to be so strongly constrained;   

\item It is not clear how  the state $|\Psi\rangle$ is to be selected. And if matter fields divide naturally into subsystems, the interpretation of semiclassical metrics derived from entangled states is far from clear; 

\item There is a stability problem for solutions of the equation arising from the fact that r.h.s is made from covariantly conserved tensors   that have higher order derivatives \cite{Ford:2005qz};

\item If the metric and state are expanded in a perturbation series around a flat (or other symmetric) background $\eta$ and a ``vacuum" state $|0\rangle$,  
\be
 g = \eta + \epsilon h^{(1)} + \epsilon^2 h^{(2)} + \dots,\ \ \ \   |\Psi \rangle = |0\rangle + \epsilon |\psi^{(1)}\rangle + \epsilon^{(2)} |\psi^2\rangle + \cdots,
\ee 
  then the leading order order term generates the ``cosmological constant" problem, and subsequent terms in the expansion are ill-defined \cite{Husain:2015dxa}.
     
\end{itemize} 

For these reasons it is  be useful to seek other formulations of general relativity coupled to quantum matter, at least for simpler systems where explicit calculations might be possible. The simple model of Newtonian gravity coupled to a quantum particle suggests a path that might be generalizable to general relativity. The Newton-Schr\"odinger equation \cite{Diosi:2014ura,Bahrami:2014gwa,Giulini:2012zu} for a single self-gravitating particle is defined by the following coupled set of equations
 \bea
 \nabla^2 \phi &=& 4\pi G\  m |\psi|^2, \\
i\hbar \frac{\p \psi}{\p t}  &=& -\frac{\hbar^2}{2m} \nabla^2 \psi - m\phi \ \psi. 
\eea
Generalization to multi-particle systems is immediate; the potential has terms that include the gravitational energy between particles, in addition to the self-energy terms for each particle.  These equations suggest an interesting interplay between the Schr\"odinger equation's tendency to spread wave functions, and gravity's tendency to localize. This may be tested numerically by starting with an initial wave function $\psi(x,0)$, computing its potential $\phi(x,0)$ via the first equation, and using  it in the second to compute the wave function at the next time step $\psi(x, \delta t)$. This process is then continued until some desired time is reached \cite{Carlip:2008zf}. These considerations motivate our generalization of the Newton-Schr\"odinger equation to cosmology.  

\subsection{Friedmann-Schrodinger equation}

Let us consider the flat homogeneous and isotropic cosmology coupled to a scalar field in an arbitrary potential. The  metric  is 
\be
ds^2 = -dt^2 + a^2(t)(dx^2 + dy^2 + dz^2),
\ee
and the Hamiltonian and energy density of the scalar field are 
\be
H = v_0\left(  \frac{p_\phi^2}{2a^3} +  a^3 V(\phi) \right) ,\ \ \ \ \ \rho = \frac{H}{v_0a^3} = \frac{p_\phi^2}{2a^6} + V(\phi),\label{classH}
\ee
where $v_0$ is a fiducial coordinate volume, $v_0a^3$ is the physical volume, and the canonical Poisson bracket of the scalar field is  $\{\phi, p_\phi \} = 1/v_0$ after symmetry reduction. (The resulting classical equations of motion are of course independent of $v_0$). The Einstein equations lead to the first and second Friedmann equations for the coupled system. 

The semiclassical system we propose  uses the usual  quantization of the  scalar field. The Hilbert space is
 ${\cal H} = L^2 (\mathbb{R}, d\phi)$ and the elementary operators are 
 \be
  \hat{\phi} \psi(\phi,t) = \phi \psi(\phi,t),  \ \ \ \ \ \  \hat{p}_\phi\psi = -\frac{i}{v_0} \frac{\p \psi}{\p \phi}.  
 \ee
 The model is defined by the set of coupled (``Friedmann-Schr\"odinger") equations: 
\bea
\left(\frac{\dot{a}}{a}\right)^2 &=& \frac{l_p^2}{3} \ \langle \psi | \hat{\rho}| \psi \rangle    + \frac{\Lambda}{3}  \label{scm1} \\
i \frac{\p\psi}{dt}  &=&  - \frac{1}{2 (v_0 a^3)}  \frac{ \p^2\psi}{\p \phi^2}  + (v_0 a^3) V(\phi) \psi, \label{scm2} 
\eea
where 
\be
\langle \psi | \hat{\rho}| \psi \rangle =   \int_{-\infty}^{\infty} d\phi \  \psi^* \left( - \frac{1}{2(v_0a^3)^2}\frac{\p^2 \psi }{\p \phi^2} +  V(\phi)\right) \psi.
 \ee
The first is the semi-classical Friedmann equation with source determined by the state $| \psi \rangle$, and the second is the Schr\"odinger equation for the scalar field Hamiltonian, with the mass determined by the physical $3$-volume $v_0a^3$.  

We note that unlike in the semiclassical equation (\ref{sc1}), the dynamics of the metric in eqns. (\ref{scm1} -- \ref{scm2}) is not driven entirely by the Einstein equations. Rather,  an important component of the dynamics is that of the state through the Schrodinger equation. 
There is no independent equation for $\ddot{a}$. This is as it should be since scalar field evolution is effectively replaced by the TDSE (\ref{scm2}); the (effective) Friedmann equation (\ref{scm1}) determines $\dot{a}$ as a function of the quantum state $\psi$, and $\ddot{a}$ may be derived, if needed, from $\dot{a}$ by differentiation.  There is therefore no independent semi-classical equation for $\ddot{a}$.   With this understanding, the pair of equations defining our system is self-consistent. 

With the standard assignments of dimension, our equations may be transformed to dimensionless variables with the re-scalings 
\be
\bar{t} =t/l_P,\ \ \bar{\phi} = l_P\phi, \ \  \bar{\Lambda} = l_P^2 \Lambda, \ \ \bar{V} = l_P^4 V, \ \ \bar{\psi} = l_P^{-1/2}\, \psi.
\ee
Dropping the bars, the dimensionless equations (with $v_0/l_p^3 =1$) are 
\bea
\left(\frac{\dot{a}}{a}\right)^2 &=& \frac{1}{3} \ \langle \psi | \hat{\rho}| \psi \rangle  + \frac{\Lambda}{3} \label{effF1} \\
i \frac{\p\psi}{dt}  &=& - \frac{1}{2 a^3}  \frac{ \p^2\psi}{\p \phi^2} + a^3 V(\phi)\psi. \label{tdse}
\eea

These equations may be  integrated numerically given an initial data set $\{ \psi_0\equiv \psi(\phi, 0) \in {\cal H}, a(0)\}$. Using (\ref{tdse}),  this data gives the wave function $\psi(\phi, \delta t)$ at the next time step, and the Friedmann equation gives $a(\delta t)$ using the initial density $\langle \psi_0| \hat{\rho} |\psi_0\rangle$. This process is continued for multiple time steps to obtain a long time evolution.    

\subsection{Comparison with the usual semi-classical approximation} 

It is illustrative to compare the above equations with the conventional semi-classical approximation for cosmology derived from (\ref{sc1}). Despite the 
issues with this approach mentioned above, it is possible in simple cases to set-up and solve these equations. Cosmology is one case in point. 
There is a version of these equations which arise in the Hamiltonian framework by choosing a Gaussian (or other) scalar field state peaked on a classical phase space point $(\phi_0, p_0)$, computing the expectation value of the Hamiltonian operator in that state to obtain an effective Hamiltonian $H^{eff}(\phi_0, p_0)$, and then deriving the semi-classical evolution from $H^{eff}$. This process replaces the TDSE in our approach. 

For the Gaussian state 
\be
\psi(\phi; p_o,\phi_0,\sigma) = N \exp\left[ -(\phi-\phi_0)^2/\sigma^2 -i \phi p_0  \right]
\ee
the effective Hamiltonian derived from the operator (\ref{classH}), for the  quadratic potential $V(\phi) = V_0\phi^2/2$, is 
\be
H^{eff} = \langle \psi | \hat {H}  |\psi  \rangle =   \frac{1}{a^3}\left(p_0^2 + \frac{1}{\sigma^2} \right) + \frac{V_0a^3}{2} \left(\phi_0^2 + \frac{\sigma^2}{4} \right).
\ee
Imposing a Poisson bracket on the peaking values  $(\phi_0, p_0)$ leads to the semi-classical equations
\bea
\dot{p}_0 &=& \left\{p_0, H^{eff} \right\}=- a^3V_0 \phi_0 \nonumber\\
 \dot{\phi}_0 &= &\left\{\phi_0 , H^{eff}\right\} = \frac{p_0}{a^3}  \\
\left( \frac{\dot{a}}{a} \right)^2 &=&  \frac{1}{3}\ \rho^{eff}(\phi_0,p_0;\sigma) + \frac{\Lambda}{3}, \label{usualSC}
\eea
where $\rho^{eff} = H^{eff}/a^3$. The first two equations are identical to the classical equations for the scalar field, but the Friedmann equation now contains the state width corrections depending on constant $\sigma$ coming from the source $\rho^{eff}$. This modifies the gravitational dynamics. The key difference from our proposed approximation is that here  there is no quantum evolution equation for the state; the state remains Gaussian of fixed width as the Universe evolves. For comparison, we present below numerical results from this approximation as well.

\subsection{The Kasner-Schr\"odinger equation}

The model we have described above may be extended to homogeneous anisotropic cosmology. Although we will not study this case numerically it is useful to see formulate the equations. A useful form of  the   metric  is 
\be
ds^2 = -dt^2 + a^2(t) \beta_{ij} dx^i dx^j,  
\ee
where the matrix $ \beta = \text{diag}( 2\beta_+  + 2\sqrt{3} \beta_-, 2\beta_-  - 2\sqrt{3} \beta_-, -4\beta_+)$  is the Misner parametrization \cite{Misner:1969hg} of anisotropies  given by the functions $\beta_\pm (t)$. 

Following a procedure similar to the above, the semiclassical equations for this model coupled to a scalar field are 
\bea
\left(\frac{\dot{a}}{a}\right)^2 &=& \frac{1}{2a^6} \left( p_+^2 + p_-^2  \right) + \frac{1}{3} \ \langle \psi | \hat{\rho}| \psi \rangle    + \Lambda, \label{effF2} \\
\dot{\beta}_\pm &=& \frac{p_\pm}{a^3}, \label{anisot}\\
i \frac{\p\psi}{dt}  &=& - \frac{1}{2  a^3}  \frac{ \p^2\psi}{\p \phi^2}  + a^3 V(\phi) \psi. 
\eea
The first is the Friedmann equation generalized to the Bianchi I case. The second are the Hamiltonian evolution equations for the anisotropies $\beta_\pm$, where $p_\pm$ are the conjugate momenta. (Since the Hamiltonian constraint does not depend on $\beta_\pm$ in the model, the momenta $p_\pm$ are constants of the motion.  The last equation is TDSE of the scalar field, and is identical to the one for the FLRW case. (All equations are written in dimensionless form as before.) 

For other Bianchi models, there  is a potential $V(\beta_-,\beta_+)$. This modifies the gravitational equations, where now $p_\pm$ are no longer constants of the motion, and the Friedmann equation acquires a curvature term coming from the potential.  
 
Initial data sets for numerical integration are now $\{a(0), \psi(0), \beta_\pm(0)\}$, together with a specification of the constant anisotropy momenta $p_\pm$.  Numerical integration proceeds in a similar way to the FLRW case, except that there are now two additional classical equations
 (\ref{anisot}) for the evolution of anisotropies. Nevertheless, quantum back reaction on the metric still effects $\beta_\pm(t)$ because the scale factor $a$ depends on the state $\psi$ through (\ref{effF2}). The extension to all Bianchi models is immediate as this involves adding the appropriate  potential $V(\beta_-,\beta_+)$ to the Hamiltonian constraint equation (\ref{effF2}) and modifying the r.h.s of (\ref{anisot}).
 
 We note in concluding this section that the semiclassical approximation we have defined is one way of coupling a classical system to a quantum one. Several other approaches, distinct from ours, have been studied. Among these, with specific application to cosmology, is one that uses a fixed quantum state to compute the expectation value of the matter Hamiltonian. This ``effective" Hamiltonian is then used in the Hamiltonian constraint to derive a modified classical dynamics\cite{Hossain:2009ru, Hassan:2014sja,Ali:2017fhp}; there is no TDSE, and the method closely resembles the conventional semiclassical approximation  (\ref{sc1}).  Another approach uses the Born-Oppenheimer approximation on a  toy quantum model that mimics gravity coupled to matter \cite{Padmanabhan:1989aa,Singh:1989aa}, with emphasis on constructing an expansion in $\hbar$; again the equations in this formulation are different from the ones we propose. A more recent paper \cite{Vachaspati:2017jtw} considers the interaction of heavy and light harmonic oscillators in perturbation theory, again with a view to computing back reaction on a classical system.

\section{Numerical Method and Results}

 We obtained numerical solutions of eqns. (\ref{scm1}) and  (\ref{scm2}) for  a variety of potentials $V(\phi)$.   As noted above, time evolution occurs in two parts -- one is the Friedmann equation, an ordinary differential equation which is integrated  using a straightforward finite-difference method. The TDSE  is relatively non-trivial to solve. A key requirement is preservation of  unitarity at every time step. To accomplish this we follow an algorithm given in ref.~\cite{van-Dijk:2007aa} which generalizes the Crank-Nicolson scheme to the Pad\'e approximation of the unitary evolution operator.  The details of the procedure are given in the Appendix \ref{Appendix:A}. The code was written in Python and executed for a variety of initial data sets and potentials. 
 
\begin{figure}[t!]
\centering
\includegraphics[scale=0.4]{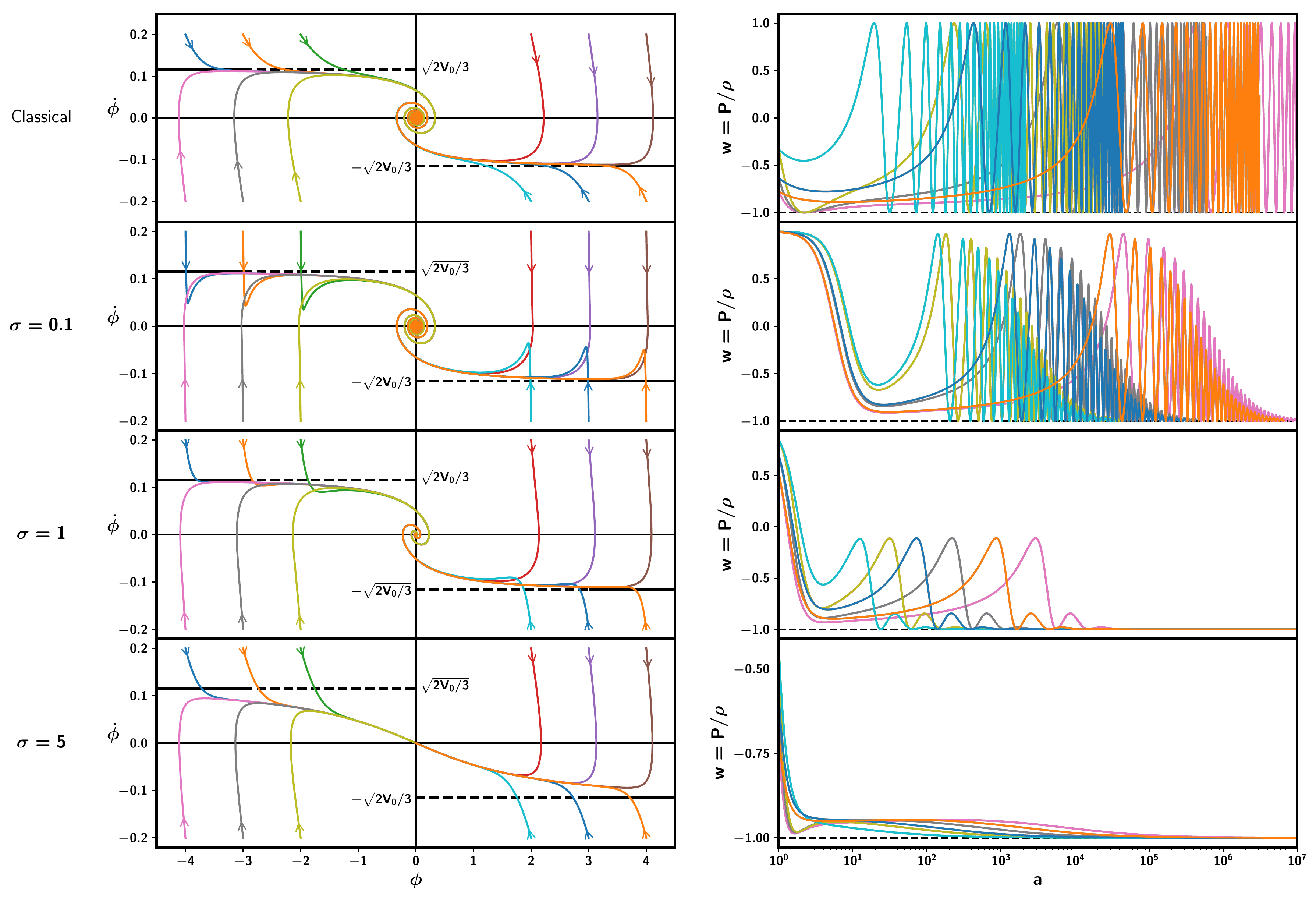}
\caption{Phase plots for the classical scalar field (top frame), followed by the those for the usual semi-classical approximation (\ref{sc1}) with potential $V=V_0\phi^2/2$ ($V_0 = 0.02$ in this case). As the width $\sigma$ of the Gaussian wave function increases, it is apparent that the damped oscillatory behaviour decreases as the field and its time derivative approach zero. In all but the classical case, inflation is eternal due to the non-vanishing expectation value of the potential.}
\label{fig:0}
\end{figure}

We present first the phase plots for the usual semi-classical approximation defined by eqns. (\ref{usualSC}) for the quadratic potential, for a range of wave function widths $\sigma$. (This is a set of ODEs that may be integrated using standard packages.) These appear in Fig. \ref{fig:0}. There are two important differences from classical theory:  (i) as $\sigma$ increases, the oscillatory exit from inflation disappears, and (ii) the equation of state settles eventually to $w=-1$ due to the presence of zero point fluctuations of the potentials (which is a function of $\sigma$), and the kinetic part of the energy density vanishes.  This gives a late time ``cosmological  constant" but its value must be tuned in the initial data to agree with the observed value today. 

Turning to our semiclassical model, the integration grid was set with $\phi \in  [-15,15]$, with initial scale factor chosen to be $a(0) =1$.  Evolution for the TDSE was implemented with a  6-by-6 Pad\'e approximation for the unitary operator, and a 13-point scheme for the second spatial derivative. This ensured probability conservation to a high degree of accuracy, with errors of order $10^{-6}$. The details of the potentials and initial quantum states, (normalized before evolution),  are as follows:

\noindent{\underbar{Quadratic:}}   $V(\phi) = V_0\,\phi^2/2$ with $V_0 =0.02, 0.009, 0.005]$ and  initial quantum state 
\be
\psi (\phi,0) = e^{-(\phi-\mu_1)^2/\sigma^2} e^{-i k_0 \phi} \sin(3\phi) + e^{-(\phi-\mu_2)^2/\sigma^2} e^{i k_0 \phi} \sin(3\phi)\nonumber
\ee  
with  $\mu_1 = 7$, $\mu_2 = -7$, $\sigma = 2$, $k_0 = 14$. The initial expectation value and fluctuation of the field are 
$\langle \phi \rangle_0 = 0, \Delta \phi_0 =7.07$.

\begin{figure}[t!]
\centering
\includegraphics[scale=0.4]{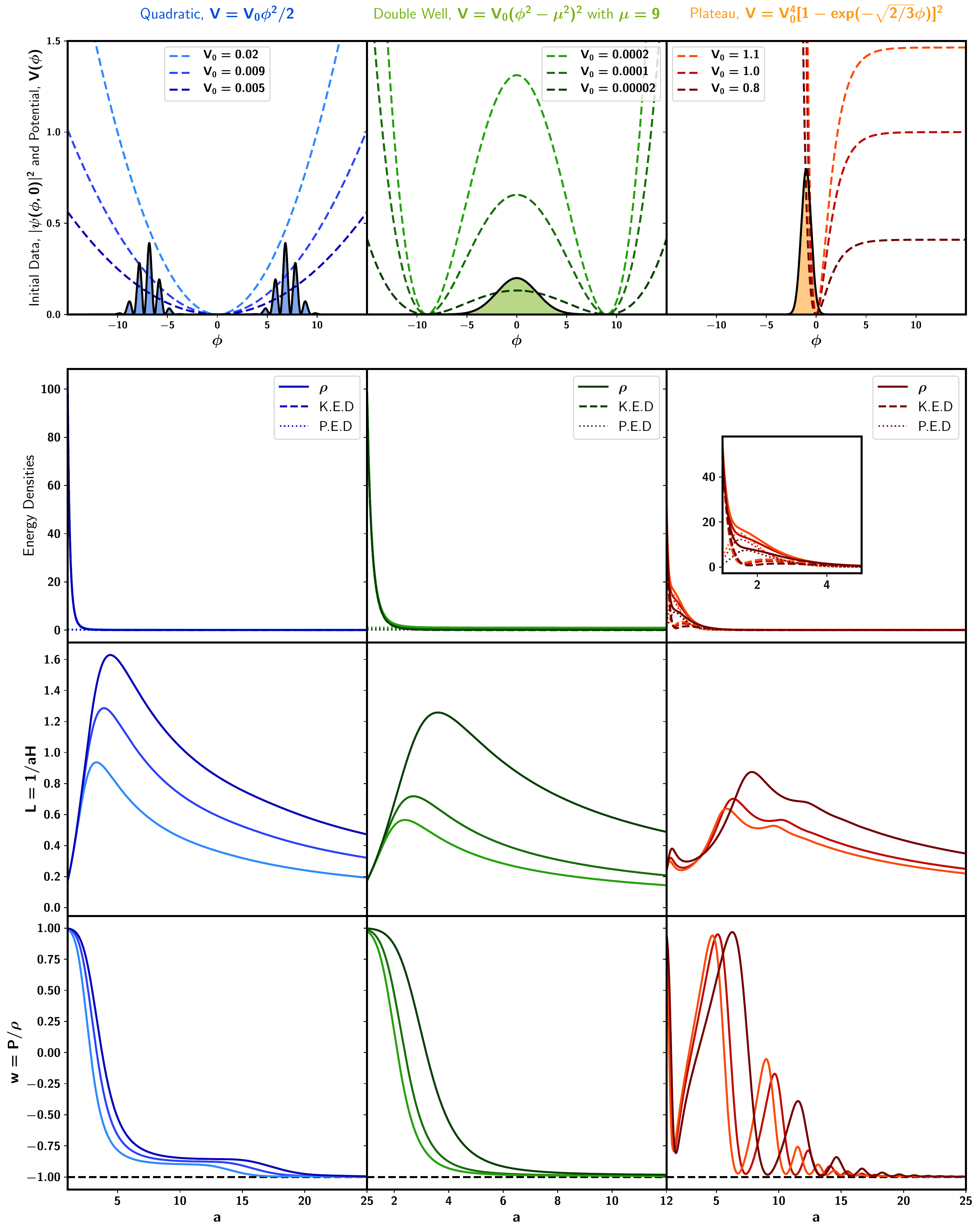}
\caption{ The figure shows  time evolution of energy densities -- total, kinetic and potential, co-moving Hubble radius, and equation of state parameter for three different initial datasets for  the potentials indicated. The common feature in all the plots is a dynamical and natural onset of an inflationary phase, regardless of initial data. Quantum fluctuations of the field give $P=-\rho$ at late times, again independent of initial state and potential.}
\label{fig:1}
\end{figure}
\noindent\underbar{Double Well:}   $V=V_0 (\phi^2 - \mu^2)^2$, $\mu = 9$ with $V_0 =0.0002, 0.0001, 0.00002$, and  
\be
\psi (\phi,0) = e^{-\phi^2/\sigma^2} e^{-i k_0 \phi}\nonumber
\ee  
with $\sigma = 4$, $k_0 = 14$. The initial expectation value and fluctuation of the field are $\langle \phi \rangle_0 =0$ 
and $\Delta \phi_0 = 2.0$. 

\noindent \underbar{Plateau:} $V(\phi) = V_0^4 \left[1- \exp (-\sqrt{2/3}\phi)\right]^2$  with $V_0 =1.1, 1.0, 0.8$ and  
\be
\psi (\phi,0) = e^{-(\phi-\mu)^2/\sigma^2} e^{-i k_0 \phi}\nonumber
\ee  
with $\mu = -1$, $\sigma = 1$ and $k_0 = 10$.  The initial expectation value and fluctuation of the field are 
$\langle \phi \rangle_0 = -1$ and $\Delta \phi_0 = 0.5$. 

\begin{figure}[t!]
\centering
\includegraphics[scale=0.7]{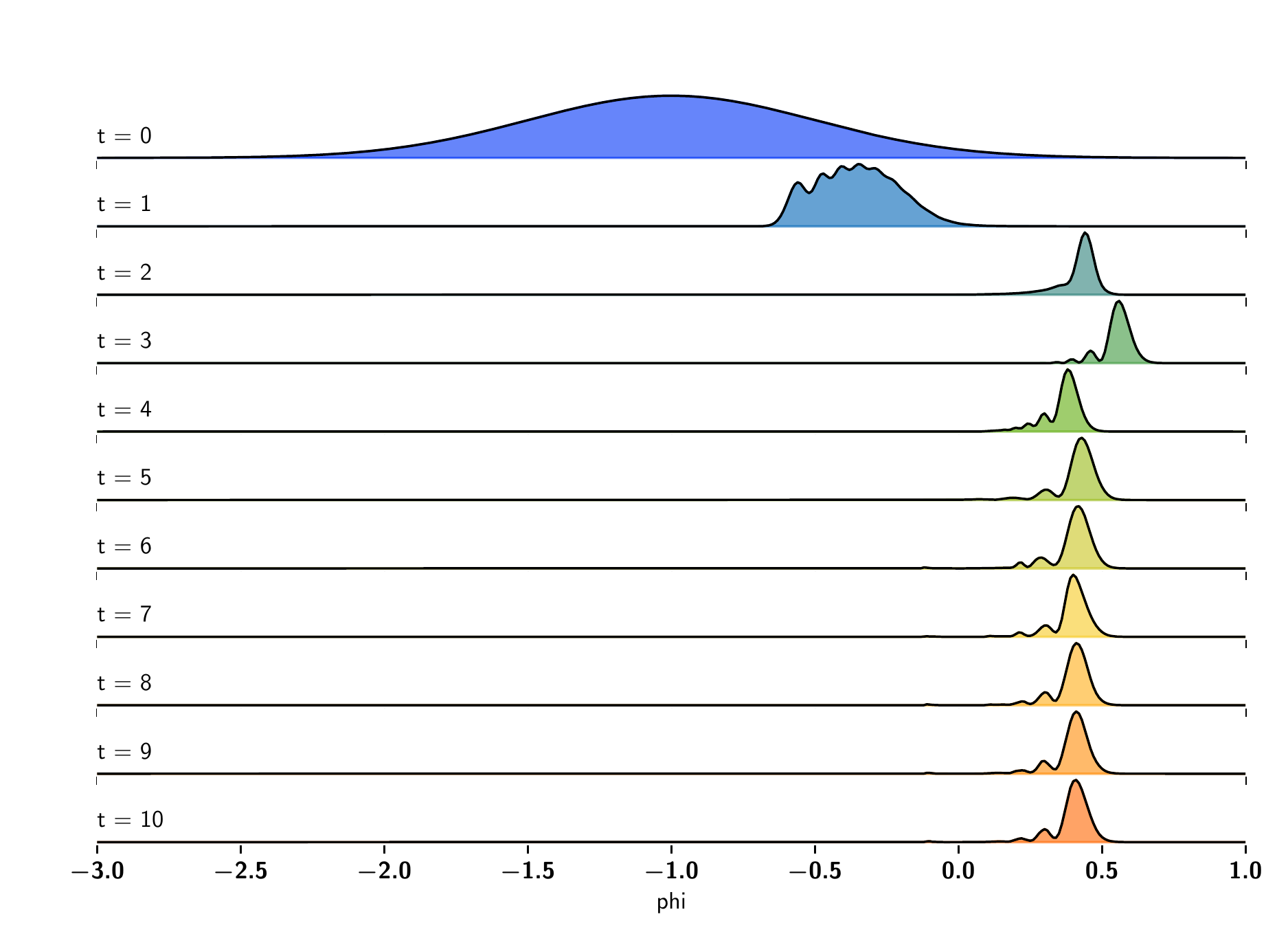}
\caption{This is the evolution of the wave function for the plateau potential; this shows an initial kinetic energy dominated phase where the packet is moving but gets frozen at late-times upon entering the potential dominated phase. (The scaling of the y-axes is different for different times and thus the height of peaks should not be compared.) }
\label{fig:2}
\end{figure}

Fig. \ref{fig:1} shows typical results for these potentials for the range of parameters indicated, and Fig. \ref{fig:2} shows the evolution of the wave function for the plateau potential.  We note the following features of these results which appear to be generic (based on many similar runs with a  variety of initial wave functions):

\begin{enumerate}

\item Evolution begins with a kinetic energy dominated phase if the initial kinetic energy density is larger than the initial potential energy density.

\item As the Universe expands, the potential squeezes due to the $a^3$ factor multiplying $V$. Evolution becomes potential dominated at late-times whenever the kinetic energy falls faster than the time it takes for the potential to squeeze in, effectively freezing the motion of the wave function.  This is apparent in Fig. \ref{fig:2} after $t=5$. 

\item The equation of state parameter $w$ dynamically approaches $-1$.

\item  The Hubble slow roll parameter, $\epsilon_H = -\dot{H}/H^2$ which is proportional to $w$ dynamically falls below unity without fine-tuning of initial data. 
\end{enumerate}

 \section{Summary and Discussion}
 
 Our main results are as follows: (i) we postulated a semiclassical approximation with back reaction for the coupled evolution of a classical FLRW cosmology and quantum scalar field via the equations  (\ref{scm1}-\ref{scm2}), and (ii) used numerical solutions of these equations (for a variety of potentials) to show that the inflation occurs generically regardless of initial quantum state or potential. 
 
 The mechanism by which this occurs is evident from our equations. As the Universe expands the ``mass"  $v_0a^3$ in the TDSE (\ref{scm2}) increases, which results in a  decreasing kinetic energy; the same factor contracts the potential. This leads dynamically to a potential dominated Universe at late times. This behaviour is independent of the form of the potential and the  initial quantum state. 
  
 In the conventional treatment, the zero-mode homogeneous isotropic sector is treated as a classical background that is not subject to back reaction. Our approach introduces a mechanism for back reaction, and at the same time treats the zero-mode quantum mechanically. This puts the zero-mode on the same footing as the inhomoegeneous ones (which alone are treated quantum mechanically in the standard approach).    
 
 There are potentially interesting consequences of these results for thinking about the inflationary paradigm and its criticisms. Prominent among these is the question of the naturalness of inflation -- how large is the class initial data that leads to inflation, and how dependent is this on the form of the potential?   As we have seen, the semi-classical approximation we have presented provides a much larger sets of quantum initial data -- square-integrable functions worth, all of which lead to inflation.  
 
One shortcoming of our equations is that  there is no natural exit from inflation due to the persistence of zero-point fluctuations of the scalar field: as the Universe expands the kinetic term in the TDSE is suppressed, the dynamics becomes potential dominated. But this inturn means that $\langle V(\hat{\phi}) \rangle$ at late times can never be zero. This gives an eternal effective cosmological constant, unless some mechanism of suppression is introduced by hand. 

Lastly, a broader point concerning the semiclassical equations we have studied is their use beyond the homogeneous and isotropic models. As indicated in Sec. II it is straightforward to introduce anisotropy. Extensions to include inhomogeneities are far more interesting, both in cosmology and in spherically symmetric gravity-scalar system. We are at present studying both these generalizations. 

\bigskip

\noindent  \underbar{Acknowledgements}  This work was supported by the Natural Science and Engineering Research Council of Canada. It is a pleasure to thank Sanjeev Seahra and Edward Wilson-Ewing for helpful discussions. 

\vfill\eject

\appendix

\section{Numerical method for the time-dependent Schr\"odinger equation}
\label{Appendix:A}

There is a subtle computational process that goes behind numerically solving the coupled Friedmann equation and the time-dependent Schr\"odinger equation (TDSE). We briefly describe the same in this section for completeness. The algorithm follows from ref.~\cite{van-Dijk:2007aa} where the authors have presented a generalization of the Crank-Nicolson (CN) method to obtain an accurate and fast numerical evolution of TDSE. The details can be found therein. 

The time evolution of a system in the Schr\"odinger picture is essentially a Unitary operation on the state as given by
\be
  \psi(x,t+\Delta t) = e^{\textstyle -iH\Delta t/\hbar}\psi(x,t).
\ee
where $H$ is the Hamiltonian of the system. The CN scheme kicks off by expanding the time-evolution operator $e^{\textstyle -iH\Delta t/\hbar}$ in a unitary approximation:
\begin{equation}\label{eq:2.4}
e^{\textstyle -iH\Delta t/\hbar} = \frac{1 - \frac{1}{2}{iH\Delta t/\hbar}}
{1 + {1\over 2}{iH\Delta t/\hbar}} + {\cal O}((\Delta t)^3).
\end{equation}
and substituting it back to obtain 
\be
  \left(1+{1\over 2}iH\Delta t/\hbar\right)\psi(x,t+\Delta t) =  
  \left(1-{1\over 2}iH\Delta t/\hbar\right)\psi(x,t).
\ee
Once the state vector $\psi(x,t)$ is discretized as $\psi_{j,n}$ on a spatial and temporal grid:
\begin{align}
&x_j = x_0 + j \Delta x,  ~~~~ i = 0,\ldots, J ~~~~ \Delta x =  (x_J - x_0)/J\nn\\
&t_n = t_0 + N \Delta t,  ~~~~ n = 0,\ldots, N ~~~~ \Delta t =  (t_N - t_0)/N\nn
\end{align}
the evolution equation becomes just a matrix equation
\begin{equation}
  A\Psi_{n+1} = A^*\Psi_n,
\end{equation}
relating the wave function at $t_{n+1}$, that is, $\Psi_{n+1}$, a column vector consisting of the $\psi_{j,n+1}$ as components with $\Psi_n$ specifying the state at the previous time step. The matrices $A$ and its complex conjugate $A^*$ specify the action of the Hamiltonian on the discretized system. A three-point finite difference formula is used usually for the kinetic term in the Hamiltonian which is a second -order spatial derivative. The discretization of the potential term is simply $V_{i,n} = V(x_j,t_n)$. With these inputs, $A$ is a tri-diagonal matrix and standard techniques can be used to solve the equation iteratively starting from a normalized initial wave function that also usually satisfies Dirichlet boundary conditions at the endpoints of the spatial grid.

The generalized CN scheme consists of two modifications in the above procedure. One is in specifying the second-order spatial derivative in terms of a more accurate $(2r+1)$-point formula
\begin{equation}
 y''(x)\equiv y^{(2)} = \frac{1}{h^2} \sum_{k=-r}^{k=r} c_k^{(r)} \,
  y(x+kh) + {\cal O}(h^{2r}),
\end{equation}
where $c_k^{(r)}$ are real constants. These constants are obtained by considering the expansions for $y(x+kh)$ and $y(x-kh)$ for $k = 1,\ldots,r$, combining them to cancel the odd derivatives and then solving the resulting $r$-equations for $r$-unknowns $y^{(2k)}(x)$. The second improvement is in time advancement where the exponential operator $\exp(-iH\Delta t)$ is replaced by the diagonal $[M/M]$ Pad\'e approximant. 
\begin{equation}\label{eq:3.1}
 e^{\textstyle z} = \frac{\displaystyle \sum_{m=0}^M a_mz^m}{\displaystyle\sum_{m'=0}^M b_{m'}z^{m'}} = \frac{a_0+a_1z+\cdots + a_Mz^M}{b_0+b_1z
+\cdots + b_Mz^M} =
\prod_{s=1}^M \left(\frac{1-z/z_s^{(M)}}
      {1+z/\bar{z}_s^{(M)}}
    \right) + {\cal O}(z^{2M+1}), 
\end{equation}
where the $a_m$ and the $b_{m'}$ are complex constants, and $z_s^{(M)}, \ s =1\dots,M$, are the roots of the numerator with $\bar{z}^{(M)}_s$ being the complex conjugate of $z^{(M)}_s$. With the following definition,
\begin{equation}
  K_s^{(M)} \equiv \frac{1 + \frac{\textstyle iH\Delta t/\hbar}{\textstyle z_s^{(M)}}}
{1 - \frac{\textstyle iH\Delta t/\hbar}{\textstyle \bar{z}_s^{(M)}}}, 
\end{equation}
we can write 
\begin{equation}
  e^{\textstyle -iH\Delta t/\hbar} = \prod_{s =1}^M K_s^{(M)} 
  + {\cal O}((\Delta t)^{2M+1}).
\end{equation}
Since $\Psi_{n+1} = e^{\textstyle -iH\Delta t/\hbar} \Psi_n$, by using the Pad\'e approximant, each such time step gets broken into $M$-substeps:
\begin{equation}
  \Psi_{n+1} = \prod_{s=1}^M K_{s}^{(M)}\Psi_n.
\end{equation}
Defining $\Psi_{n+s/M} \equiv K_s^{(M)}\Psi_{n+(s-1)/M}$, we can solve for
$\Psi_{n+1}$ recursively, starting with 
\begin{equation}
 \Psi_{n+1/M} = K_1^{(M)}\Psi_n. 
\end{equation}  
using the usual CN procedure.
  
\bibliography{Fried-Sch}

\end{document}